\documentclass[fleqn,usenatbib]{mnras}
\usepackage{newtxtext,newtxmath}

\usepackage[T1]{fontenc}
\usepackage{ae,aecompl}

\usepackage{graphicx}	
\usepackage{amsmath}	
\usepackage{amssymb}    
\usepackage{xcolor}
\usepackage{subfigure}
\usepackage{etoolbox}
\makeatletter
\patchcmd\@combinedblfloats{\box\@outputbox}{\unvbox\@outputbox}{}{%
   \errmessage{\noexpand\@combinedblfloats could not be patched}%
}%
 \makeatother

\title{Uncooled Microbolometer Arrays for Ground Based Astronomy}

\author[M. F. Rashman et al.]{
M. F. Rashman,$^{1}$\thanks{E-mail: m.f.rashman@2016.ljmu.ac.uk}
I. A. Steele,$^{1}$
S. D. Bates,$^{1}$
D. Copley$^{1}$
and S. N. Longmore$^{1}$
\\
$^{1}$Astrophysics Research Institute, Liverpool John Moores University, Liverpool, United Kingdom\\
}

\date{Accepted XXX. Received YYY; in original form ZZZ}

\pubyear{2019}

\begin{document}
\label{firstpage}
\pagerange{\pageref{firstpage}--\pageref{lastpage}}
\maketitle

\begin{abstract}
We describe the design and commissioning of a simple prototype, low-cost 10$\mu$m imaging instrument. The system is built using commercially available components including an uncooled microbolometer array as a detector. The incorporation of adjustable germanium reimaging optics rescale the image to the appropriate plate scale for the 2-m diameter Liverpool Telescope. From observations of bright solar system and stellar sources, we demonstrate a plate scale of 0.75$^{\prime\prime}$ per pixel and confirm the optical design allows diffraction limited imaging. We record a $\sim$ 10$\%$ photometric stability due to sky variability. We measure a $3 \sigma$ sensitivity of $7 \times 10^{3}$ Jy for a single, $\sim$ 0.11 second exposure. This corresponds to a sensitivity limit of $3 \times 10^{2}$ Jy for a 60 second total integration. We present an example science case from observations of the 2019 Jan total lunar eclipse and show that the system can detect and measure the anomalous cooling rate associated with the features Bellot and Langrenus during eclipse.
\end{abstract}



\section{Introduction}
In this paper, we describe the design and commissioning of a simple prototype low-cost mid-infrared (mid-IR) instrument, built using commercially available components and an uncooled microbolometer array as a detector.

Ground based mid-IR ($\sim$5-20 $\mu$m) astronomical observing is very challenging and has been traditionally viewed as impossible for simple, low cost instruments due to the very high thermal background and the requirement for specialist detector systems. Quantitative mid-IR astronomy began in the 1960's \citep[e.g.][]{Low61} using heavily cooled single element bolometric detectors on small telescopes. Since the 2000's high sensitivity, {multi-pixel detectors} are in operation at 8-m class facilities such as Gemini (Michelle; \citealt{Gla97}, T-Recs; \citealt{Tel98}), VLT (VISIR; \citealt{Lag20}, MIDI; \citealt{Lei03}), and GTC (CanariCam; \citealt{Pack05}). Such instruments are highly complex and the technologies are not easily adapted to smaller 1-2m class telescopes due to their extensive cooling systems and high detector cost.

Since the 1980's many advances have been made in the development of mid-IR uncooled microbolometer arrays for defence, security and industrial applications. These detectors use vanadium oxide (VOx) microbolometers as the focal plane arrays (FPA) which can deliver noise-equivalent temperature differential (NEdT) measurements of $<$200mK \citep{Beni17} under factory conditions, and are available integrated into commercial camera systems for $<$10000 GBP by manufacturers such as FLIR. In laboratory testing we have applied analogous, standard astronomical instrumentation techniques to characterise the random and spatial noise present in uncooled microbolometers systems. We have shown that {the noise properties of these focal plane arrays are dominated by fixed pattern noise which varies on timescales of $<$ 0.5 seconds \citep{Rash18}}. This can limit operational NEdT to $>$ 50mK  and has the potential to restrict the use of these systems in astronomy. Apart from observations of the moon with very small ($<$200mm) telescopes \citep[e.g.][]{Shaw15,Vol12}, they have never been tested in ground based applications, although they have been used in some high altitude experiments \citep{Tsang15} where thermal background noise is naturally much lower.
\begin{figure*}
	\includegraphics[width=\textwidth]{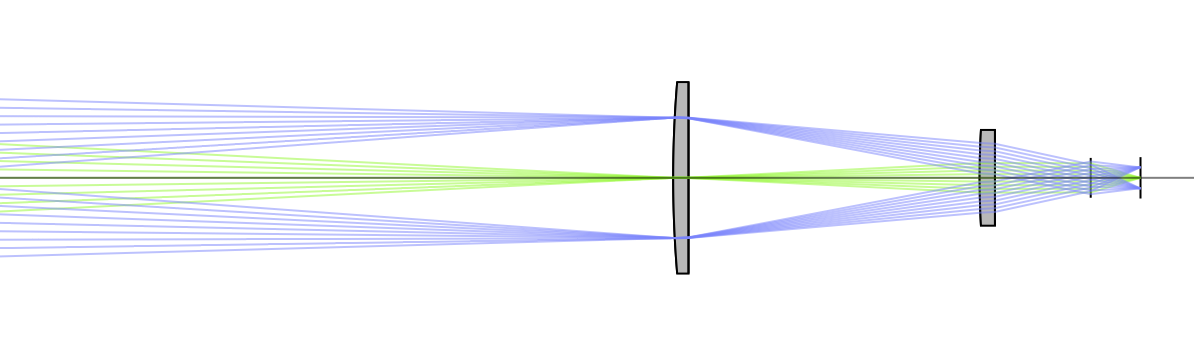}
    \caption{Side view ray trace for on- and off- axis beams of 2.7 arcmin. The 50mm diameter field and 25mm diameter collimator lens bend the rays incident off the secondary mirror onto the camera system.}
    \label{fig:ray_trace}
\end{figure*}
\begin{table*}
    \centering
    \caption{Combined optical prescription of the LT and Prototype. All dimensions are in mm.}
	\label{tab:prescription}
	\begin{tabular}{llllll} 
		\hline
		Comment & Type & Radius of Curvature & Thickness & Glass & Semidiameter\\
		\hline
        Source & Standard & inf & inf & & 0.000\\
        Primary & Asphere & -12000.0000 & 0.0000 & MIRROR & 1000.0000\\
        & Standard & inf & -4315.385 & & 1000.0000\\
        Secondary & Asphere & -4813.0000 & 0.0000 & MIRROR & 308.0000\\
        & Standard & inf & 5615.889 & & 308.0000\\
        Field Lens 50mm & Standard & 300.3900 & 4.000 & GERMANIUM & 25.000\\
        & Standard & inf & 76.000 & & 25.000\\
        Collimator 25mm & Standard & 225.0000 & 4.000 & GERMANIUM & 12.500\\
        & Standard & inf & 25.000 & & 12.500\\
        Lens & Paraxial & inf & 13.000 & & 5.200\\
        Detector & Standard & inf & 0.000 & & 5.400\\
		\hline

	\end{tabular}
\end{table*}

{More readily available mid-IR observing resources could have significant scientific impact in time domain astrophysics, where carrying out any kind of monitoring programme of variable sources at such wavelengths is currently impossible. Examples of objects where mid-IR observing is of particular value includes Blazars (where emission of the jet and torus can be traced at 10 $\mu$m), dust forming novae (DQ Her objects) and infrared variables (e.g VISTA VVV sources). However, uncooled systems such as the one described in this paper, are limited to observations of very bright (several hundred Jy) objects and are best suited to monitoring of solar system objects (comets, asteroids) and galactic objects. While this technology will never be competitive with the cooled technologies employed at the major, large facilities, successful use on 1-2m class telescopes would provide the opportunity to expand at low cost the availability of mid-IR observing of these bright objects. In addition greater availability of mid-IR observing facilities on smaller telescopes could provide a key resource to train students and early career researchers on mid-IR observing techniques and data analysis.}

In this paper we present our prototype: a small, {uncooled}, N-band ($\sim 10\mu $m) instrument, constructed from `off the shelf' components. We provide the results of a week long programme of observations conducted to test the system sensitivity and stability, and determine the feasibility of using this technology in `facility' class instruments for small telescopes. As an example science case we present observations of the cooling of lunar surface features during the 2019 January lunar eclipse.

\section{Optical Design}

The prototype was initially designed for use on the Liverpool Telescope (LT) \citep{Ste04}. However, the nature of the adjustable reimaging optics allows this design to be adapted for other 1-2m class telescopes with a similar focal ratio. The LT is an f/10 telescope with a Ritchey-Chr\'etien design. It is comprised of a 2m concave primary mirror, a 0.65m convex secondary mirror and a 0.2m science fold mirror at a 45$^\circ$ angle at the Cassegrain focus to direct light onto detectors. The LT has a 10 $\mu$m diffraction limit of 1.23 arcsec.

The prototype N-band instrument was built around a commercial mid-IR imager, produced by the manufacturer FLIR. This pre-assembled system is comprised of their Tau 2 core; a vanadium oxide microbolometer (640 x 512 pixels of dimensions 17 x 17 $\mu$m ), and a 13mm focal length lens of unspecified prescription. To rescale the image to an appropriate plate scale for use on a telescope, the 13mm focal length lens and detector were modelled as a paraxial system and ray tracing analysis for on- and off- axis light was conducted. This optical design can be seen in Fig. \ref{fig:ray_trace}. The optical prescription of the LT and prototype can be seen in Table \ref{tab:prescription}.

This prescription was then translated into a low-cost, `off the shelf' system comprised of an Edmund Optics 50mm diameter, 100mm focal length, 8-12 $\mu$m, AR coated, germanium plano-convex field lens and a Thorlabs 25mm diameter, 75mm focal length, 7-12 $\mu$m, AR coated, germanium plano-convex collimator lens. Both lenses act to collimate the light collected with the LT onto the FLIR imaging system. Both lenses have an anti-reflection coating which reduces reflection losses to $<$ 3$\%$ in the 8-12$\mu$m wavelength range. When mounted to the LT, the prototype has a plate scale of 0.75$^{\prime\prime}$ per pixel, with two pixels sampling the 10 micron diffraction limit of the LT. In practice we found that, as there was uncertainty in the location of the pupil and the presence of an aperture stop within the FLIR system, the imaging field of view (FOV) reduced to a circular aperture of 500 pixel diameter. This results in a $\sim$ 6.25$^{\prime}$ diameter FOV at LT.

\begin{figure*}

	\includegraphics[width=\textwidth]{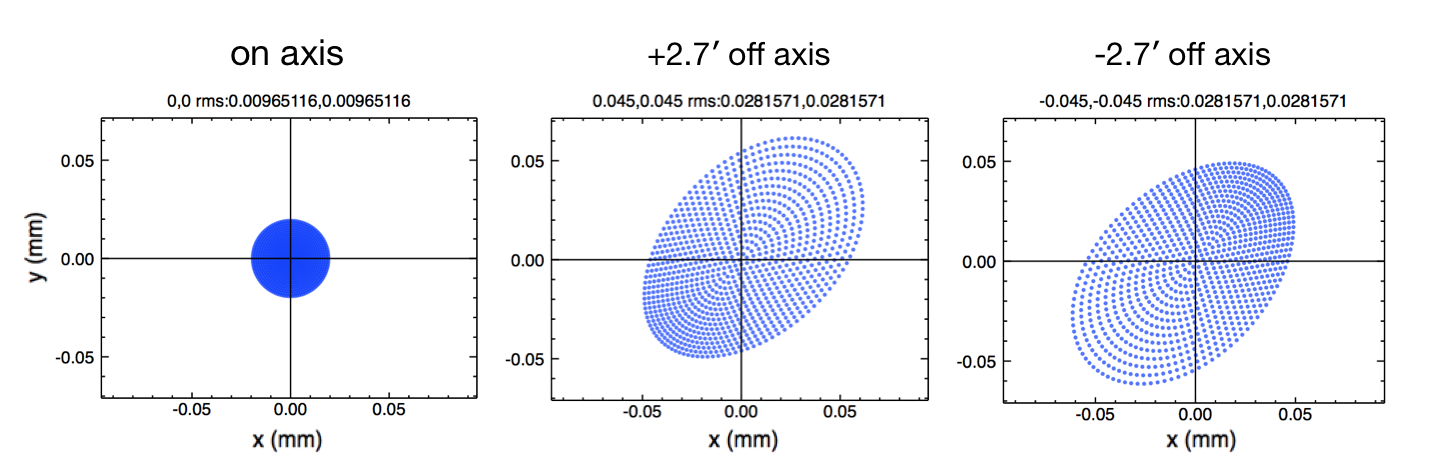}
    \caption{Spot diagrams generated assuming a 10$\mu$m wavelength point source, with no atmospheric seeing, on-axis and $\pm2.7'$ off axis. x- and y- axis are mm in the focal plane, where 0.11mm $\sim$ 4.4 arcsec.}
    \label{fig:spot}
\end{figure*}
\begin{figure}

	\includegraphics[width=\columnwidth]{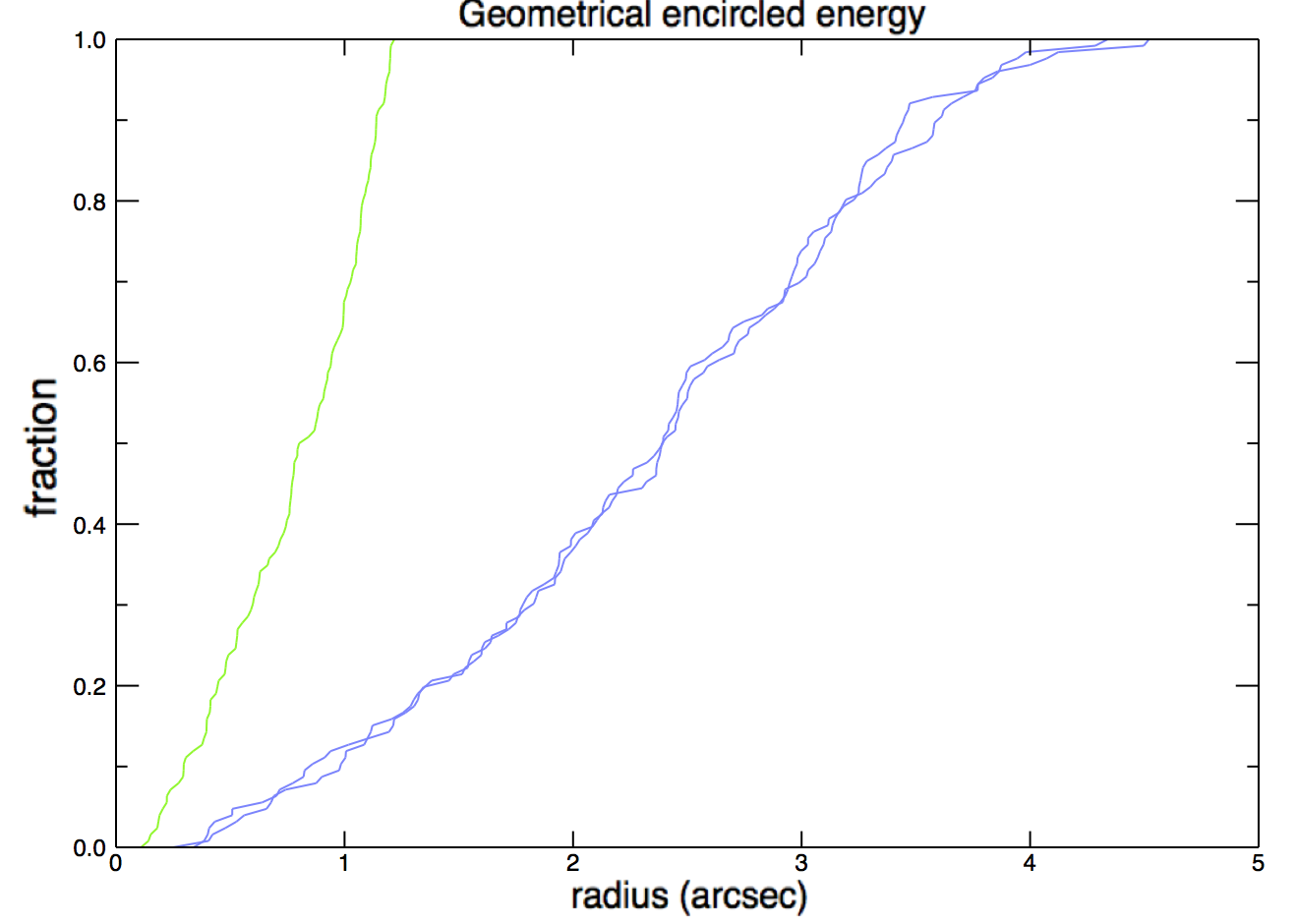}
    \caption{The geometrical encircled energy for, from left to right, on-axis and $\pm 2'$ off-axis rays for a wavelength of 10$\mu$m.}
    \label{fig:gee}
\end{figure}

Computed spot diagrams for on-axis and $\pm$ 2.7$^{\prime}$ off-axis rays are shown in Fig. \ref{fig:spot}. These figures highlight the off-axis aberration that is inherent in the optical design. The prototype has an 80\% geometric encircled energy (GEE) diameter of $<$1.1 arcsec for on-axis rays and $<$3.1 arcsec for off-axis rays, as seen in figure \ref{fig:gee}. Assuming a 2D Gaussian profile, this corresponds to a spatial full width half maximum (FWHM) of $<$ 0.7 and $<$ 2.0 arcsec for on- and off-axis rays respectively. Two pixels was therefore predicted to sample the diffraction limit for on-axis rays, with slightly worse performance off-axis.
\section{Mechanical Design}

The overall aim of this project was to build a low cost mid-IR detector that could be constructed from commercially available components. As a result of this, we opted to translate the optical design into a lens system using tubing purchased from Thorlabs. {The field lens and collimator were housed in an anodised aluminium tube of length 114mm and diameter 50mm}. The field lens was secured at the front of the lens system. The collimator was secured 76mm behind the field lens inside the tubing, using a {25mm to 50mm} adjustable adaptor to allow small changes to the focus of the lens. Two custom pieces milled from low-grade aluminium were commissioned. These pieces acted to secure placement of the FLIR 13mm lens, $\sim$ 25mm from the collimator and to attach the entire prototype $\sim$ 75mm from the telescope mounting flange, roughly at the telescope focal point. When constructed, the prototype has dimensions of $\sim$ 171mm x 60mm and a weight of 0.33$\pm$0.05 kg. This makes the prototype a very compact system and suitable for mounting on 1-2m class telescopes with a sufficient counter-balance.

The LT is a fully robotic system and therefore remote access to the instrument was essential. The ThermalCapture Grabber USB OEM was installed on the back of the FLIR system, which, paired with a Beelink J45 Mini PC running Xubuntu, allowed for remote control of data acquisition. Software was designed to collect data by downloading the image at a frequency of 9Hz whilst simultaneously displaying a live feed of observations with a delay of $<$ 1 second. This was beneficial when deploying the prototype to be able to determine parameters such as the optimum focus of the secondary mirror and telescope pointing during operation. A live feed also allowed for manual nodding of the telescope which was important for centering calibration sources and mapping non-sidereal, non-point sources (i.e. the moon).

\section{Calibration} \label{calibration}
\begin{figure*}
    \centering
    \includegraphics[width=\textwidth]{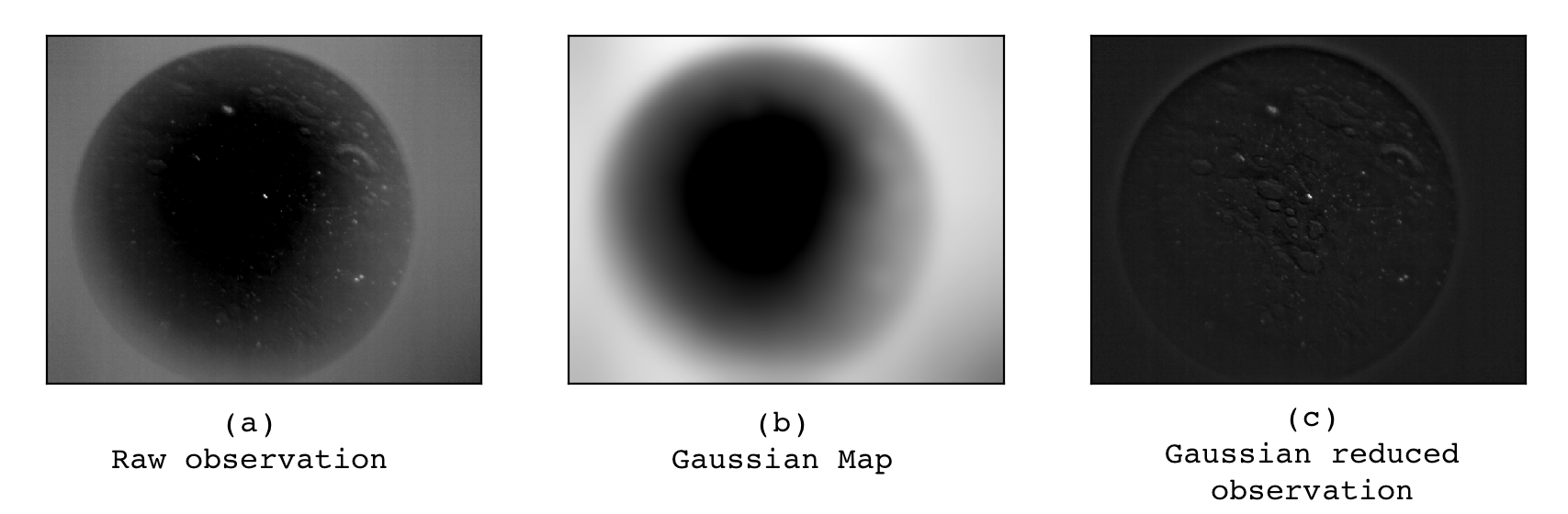}
    \caption{Observations were reduced using Gaussian Maps to remove the significant FPN present in the system. (a) is a raw observation of the eclipsing moon, (b) is the Gaussian map produced from this exposure with $\sigma=(15,15)$ pixels, and (c) is the resultant exposure reduced by applying equation \protect\ref{eq:Gaussian}.}
    \label{fig:Gaussian_reduced}
\end{figure*}
The FLIR system self calibrates by introducing an opaque shutter between the 13mm lens and the detector. This flat field calibration acts to correct any non-uniform changes in response across the detector FPA during operation. The shutter deploys at a pre-defined period of $\sim$ 2.5 minutes, but can also be triggered by any significant changes in environment or detector temperature. By applying this correction, the offset correction factors of the microbolometer elements are reset to factory standard to reduce thermal drift, which causes {counts} to increase steadily until the camera self calibrates.

Read out {values $(x,y)_{T}$ are pre-scaled by the system to have a linear relationship with temperature $T$ in Kelvins, where:
\begin{equation}
    T = (x,y)_{T} \times 0.04
    \label{eq:tlinear}
\end{equation}
To approximate these values as un-scaled counts we have assumed the following relation:
\begin{equation}
    \text{counts} = (x,y)_{T}^4
\end{equation} 
Once approximated as counts}, a scaling factor of 1.85 is introduced to correct for the f/ratio difference between the raw imaging system using the supplied camera lens, and that via the optical system.

Mid-IR background can be several orders of magnitude brighter than most astronomical sources \citep{Pie19} {and arises primarily from thermal emission from the sky, telescope and any structures visible to the detector (e.g.the optics and electronics). Temperature fluctuations in these components gives rise to noise and considerable background variability.} Sky emission (and transmission) has by far the highest spatial and temporal variability, with variations occurring on subsecond timescales. Telescopic emissions are slightly more stable, with variations occurring on timescales of tens of seconds to minutes \citep{Mas08}. IR optimised telescopes deploy chopping/nodding mechanisms frequently during observations to characterise and remove atmospheric and telescopic contributions. For the purpose of testing the prototype, and due to the LT not being optimised for IR observations, we were unable to deploy a chopping/nodding regime to remove sky contributions. Sky flats were taken prior to, and subtracted off, observations of point sources. However, for extended sources (i.e. the moon), background reduction occurred post-observation, during the data reduction stage.

Observations taken with uncooled microbolometer systems are generally dominated by fixed pattern noise (FPN). FPN is spatial noise that is generally static over short timescales, and arises from detector imperfections and variations in the responsivity, gain and noise of FPA elements. \citep{Rash18}. FPN results in a spatially heterogeneous response across the FOV. Getting a measure of the FPN present at any given time is difficult as FLIR systems have a limited range of viewable scene temperature and therefore are unable to be calibrated using exposures of super cooled surfaces. {Without the means to correct for FPN, Gaussian maps were used to `flat field' observations to correct for the large scale variations across the detector (e.g. arising from vignetting or spatially-heterogeneous variation in sensitivity) but not for the effects of FPN.} For each exposure S$_{x,y}$, a Gaussian map was created by applying a multi-dimensional, Gaussian filter with $\sigma=(15,15)$ pixels. The resulting Gaussian map g$_{x,y}$, was used to reduce the exposure, prior to data analysis, as follows: 
\begin{equation}
    I_{x,y} =\frac{S_{x,y}}{g_{x,y}} \cdot  \bar{S}_{i,j}
    \label{eq:Gaussian}
\end{equation}
where I$_{x,y}$ is the corrected exposure and $\bar{S}_{i,j}$ is the mean value inside the imaging aperture. The mean of the entire frame $\bar{S}_{x,y}$ is not used to re-scale values as the outer aperture contains no sky signal and is likely to skew values towards telescopic background. Figure \ref{fig:Gaussian_reduced} illustrates this process; (a) is a raw observation of the eclipsing moon. There is considerable FPN structure present in the FOV that can be seen as large light and dark regions. This type of FPN is characteristic of that found in all our un-reduced data. (b) is the unique Gaussian map created from (a). (c) is exposure (a) after it has been fully reduced by applying equation \ref{eq:Gaussian}.

\section{Commissioning and On Sky Testing}
The prototype was installed on the LT on 2019 Jan 19. It was deployed over three nights, including during the Lunar eclipse.
\subsection{Photometric accuracy and stability}
\begin{figure*}
    \begin{center}
        \subfigure[Mars]{%
            \label{fig:mars}
            \includegraphics[width=0.4\textwidth]{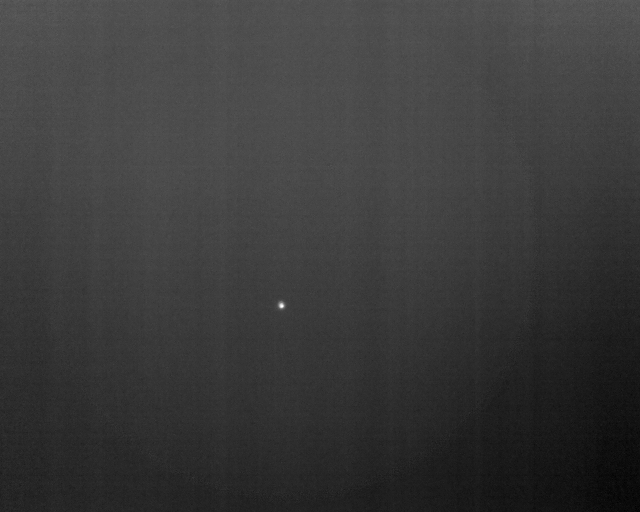}
        }%
        \subfigure[IRC+10216]{%
            \label{fig:cw}
            \includegraphics[width=0.4\textwidth]{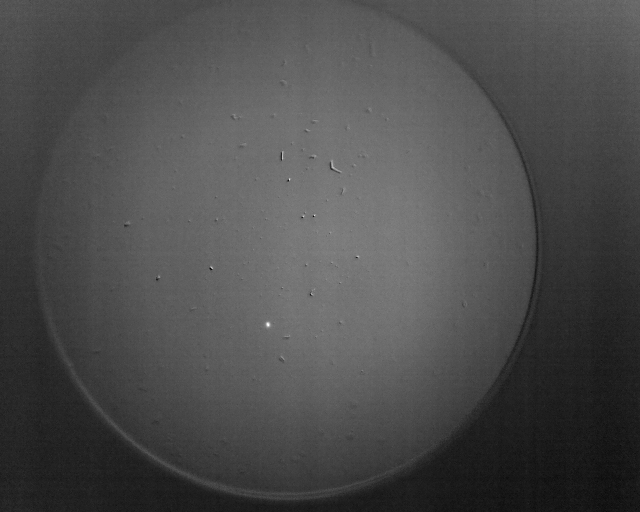}
        }%
        
    \end{center}
    \caption{Observations of (a) Mars and (b) IRC+10216 taken on 2019 Jan 22. Both (a) and (b) are median stacks created from 1000 non-Gaussian corrected, flat subtracted observations.}
    \label{fig:obs_m_cw}
\end{figure*}
\begin{figure}
    \includegraphics[width=\columnwidth]{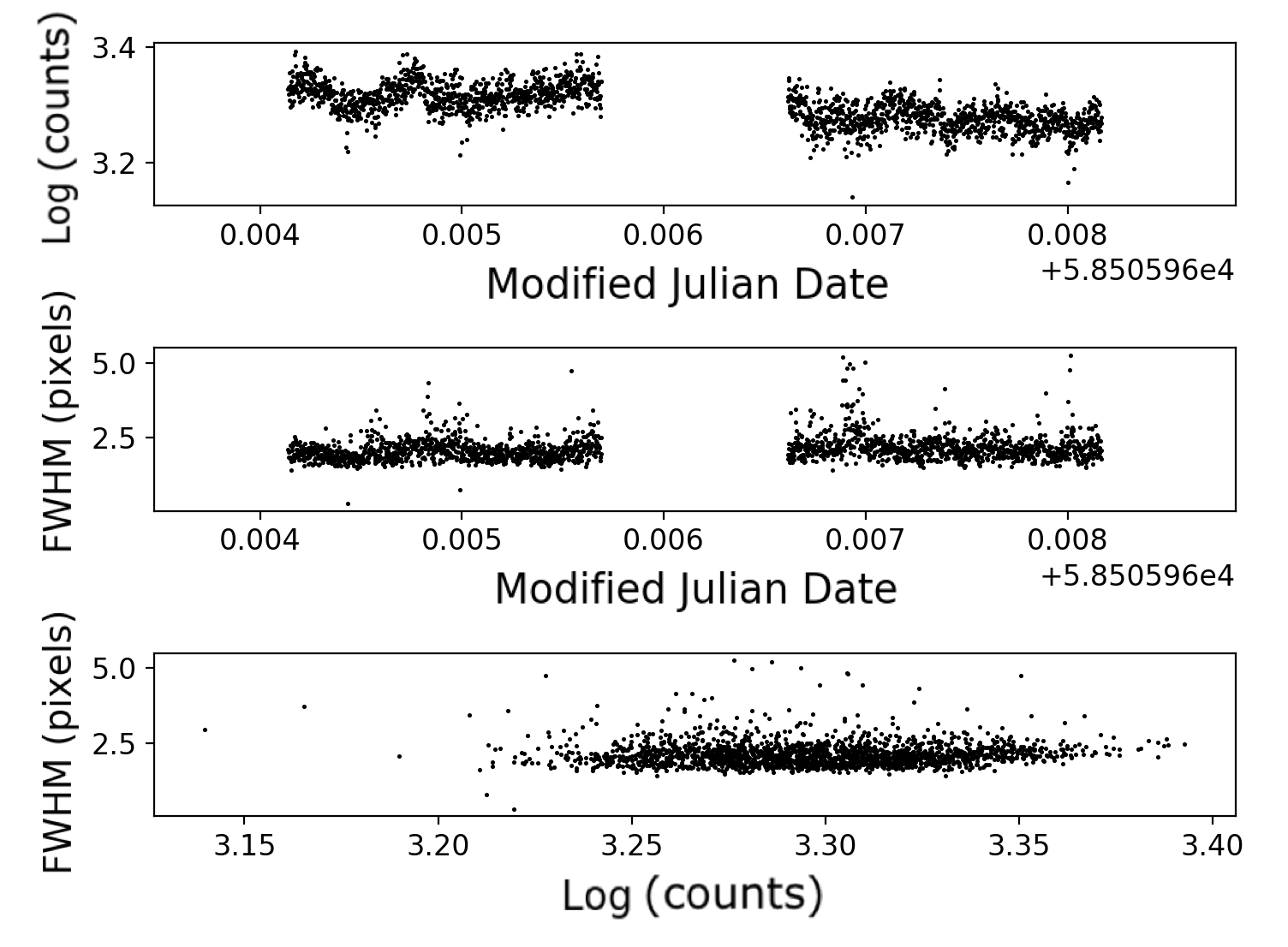}

        \caption{The variation in background-subtracted counts and FWHM for observations of IRC+10216 taken on 2019 Jan 22. Observations appear reasonably stable, with $\sim 10\%$ variability (determined from the RMS of the sample) attributed to the highly variable seeing conditions on this date, although there is no formal correlation between FWHM and counts (with a Spearman rank coefficient value of -0.03).}
    \label{fig:phot_stab}
\end{figure}

To determine the photometric performance of the system, observations were taken of two bright mid-IR sources; Mars and IRC+10216 \citep{Neu69} (see figure \ref{fig:obs_m_cw}). Offsetting the telescope pointing during our observations of Mars confirmed a pixel scale of 0.75$^{\prime\prime}$ per pixel for the system. IRC+10216 was observed at a full width half maximum (FWHM) of $\sim$ 2 pixels confirming the optical design allows diffracted limited imaging. Figure \ref{fig:phot_stab} shows the variation in FWHM and counts of IRC+10216 over a short period of 348 seconds, {taken shortly after a flat field correction to reduce the effect of thermal drift.} The $\sim 10\%$ variability in these observations can be attributed to the sky variations during observing and represent a basic estimate of the system stability.

The system accuracy was tested by comparing known and observed values of 12 micron flux for Mars and IRC+10216. A 12 micron flux of 4.75$\times 10^{4}$ Jy was obtained for IRC+10216 from the IRAS catalogue of point sources \citep{iras98}. IRC+10216 is a known variable source however for this purpose, we approximated 12 micron flux as constant. The apparent brightness of Mars depends on the sub-Earth longitude of the illuminated disk \citep{MALL07} and therefore varies seasonally and as a function of viewing angle. To determine an approximate value of 12 micron flux on our observing date for Mars, we obtained brightness temperatures at 12 and 450 microns from models derived by \cite{Wri76}, and implemented in the FLUXES routine developed for JCMT \citep{Dem13}. These brightness temperatures were then used to calculate a 12 micron flux of 76147.5 Jy using the following equation:
\begin{equation}
    S_{\nu} = \frac{2h\nu^{3}}{c^{2}}\frac{\Omega_{p}}{exp(\frac{h\nu}{kT_{\nu}})-1}
\end{equation}
where $\nu$ is the given frequency in GHz, $S_{\nu}$ is the integrated flux density of Mars, $\Omega_{p}$ is the solid angle subtended by Mars from Earth, $T_{\nu}$ is the brightness temperature and $h$, $k$ and $c$ are the Planck constant, Boltzmann constant and speed of light respectively. The flux ratio of IRC+10216/Mars for catalog and observed values were calculated as 0.62 and 0.72 respectively. {These values are consistent with a 15$\%$ uncertainty on photometric accuracy with the caveat that both sources are known to be variable.}
\subsection{Sensitivity}
Our observations of IRC+10216 were used to determine the sensitivity of the system. {Measurements of the object were obtained using apertures with a 1.5 pixel radius and sky background counts were estimated for subtraction using an annulus of 1.5-2 pixel radius. These apertures were selected using a curve of growth technique to maximise SNR}. Over a short duration ($\sim 10$ seconds) of exposures, which appear to be unaffected by sky variation, IRC+10216 has SNR=21. Assuming our observations are sky noise dominated, then SNR will be proportional to source flux. For a single exposure (with exposure time t $\sim 1/9$ seconds), a $3\sigma$ detection would correspond to a background-subtracted flux 7 times fainter than those observed in figure \ref{fig:phot_stab}, i.e. $\sim 7 \times 10^{3}$ Jy. In theory, the sensitivity can be improved by stacking exposures, with a 60 second exposure stack having a predicted $3\sigma$ detection for a source {163 times fainter than IRC+10216}, i.e. $\sim 3 \times 10^{2}$ Jy. In practice, without a nodding/chopping system, this limit is unlikely to be reached with the current setup.

\subsection{Lunar Eclipse Observation}
\begin{figure*}
     \begin{center}
        \subfigure[Penumbral eclipse]{%
            \label{fig:first}
            \includegraphics[width=0.4\textwidth]{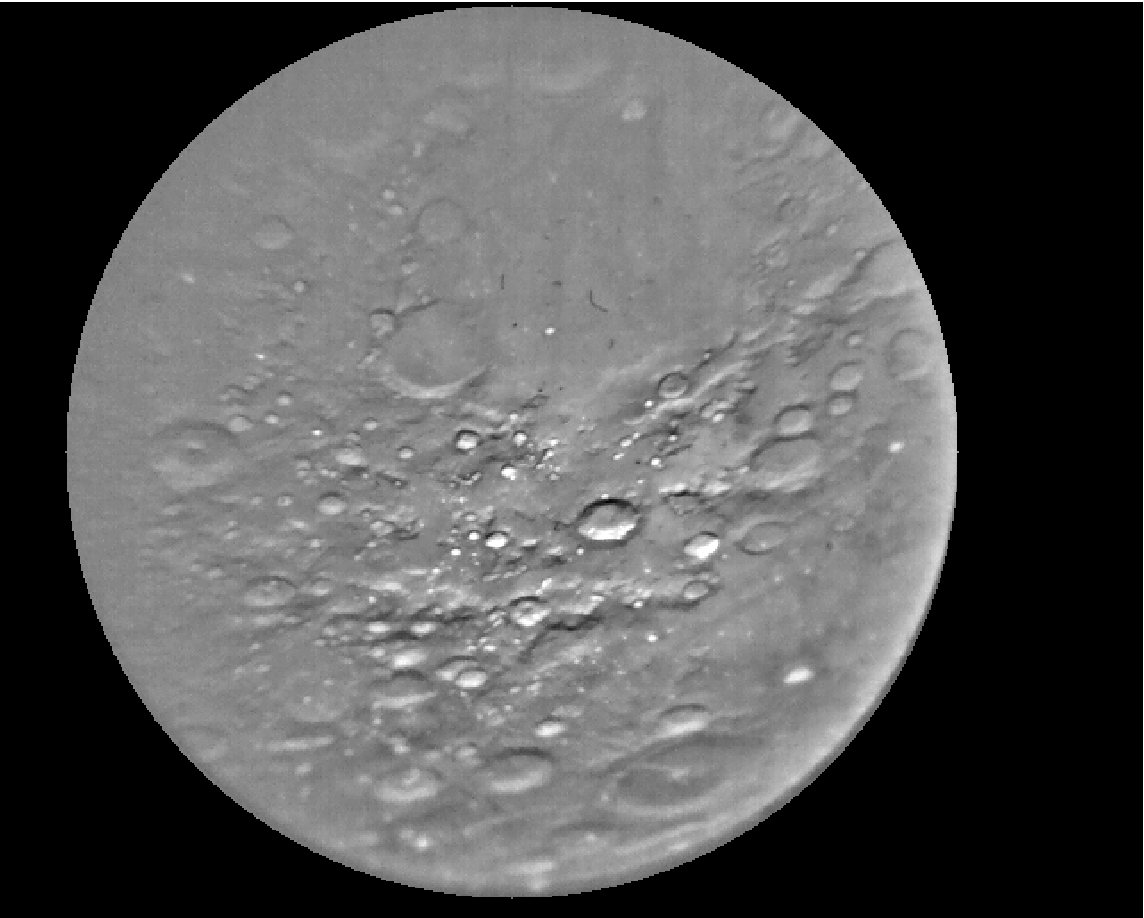}
        }%
        \subfigure[Partial eclipse]{%
           \label{fig:second}
           \includegraphics[width=0.4\textwidth]{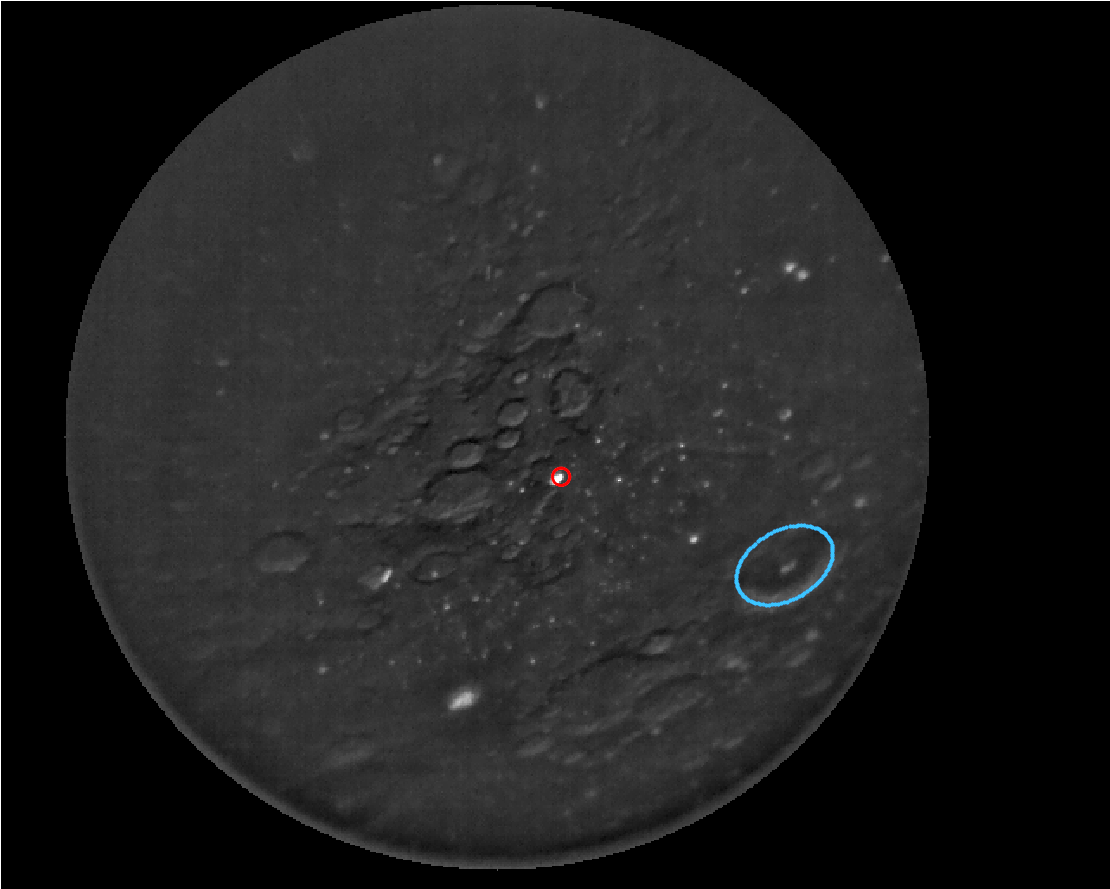}
        }\\ 
        \subfigure[Maximum eclipse]{%
            \label{fig:third}
            \includegraphics[width=0.4\textwidth]{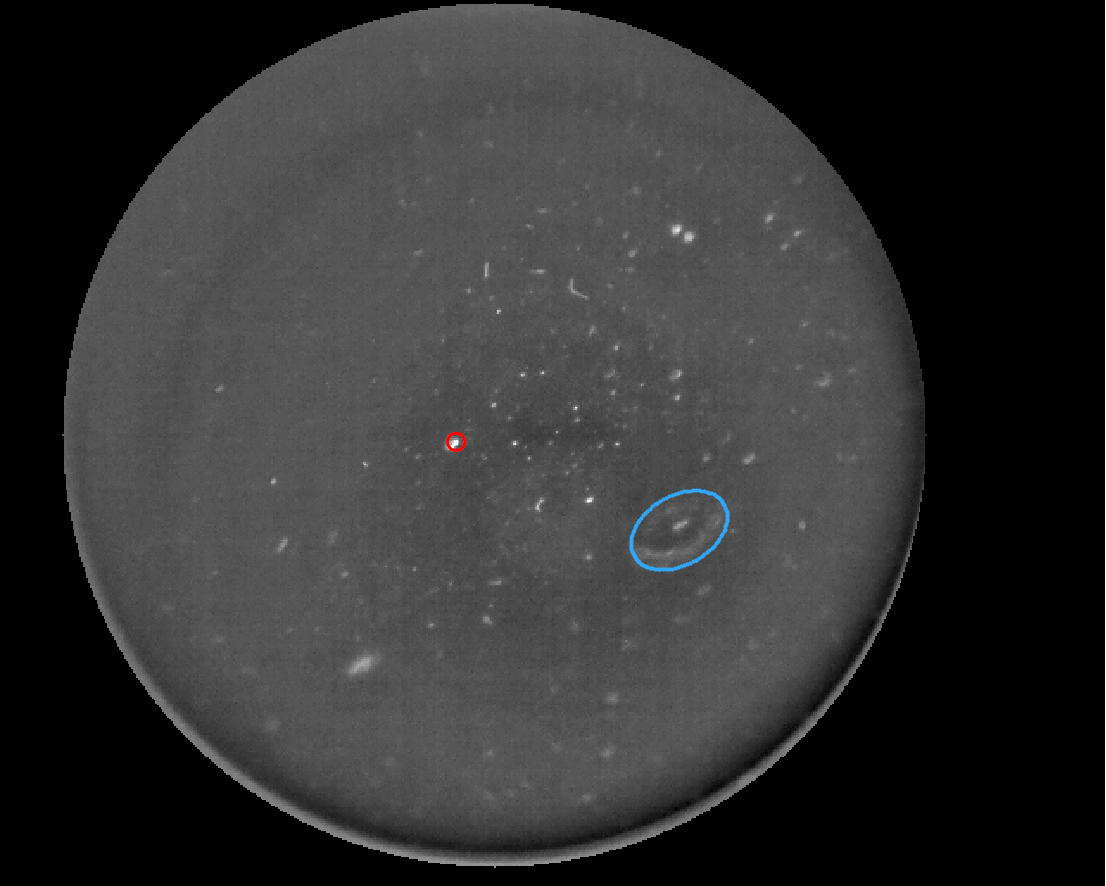}
        }%
        \subfigure[Follow up on 22/01]{%
            \label{fig:fourth}
            \includegraphics[width=0.4\textwidth]{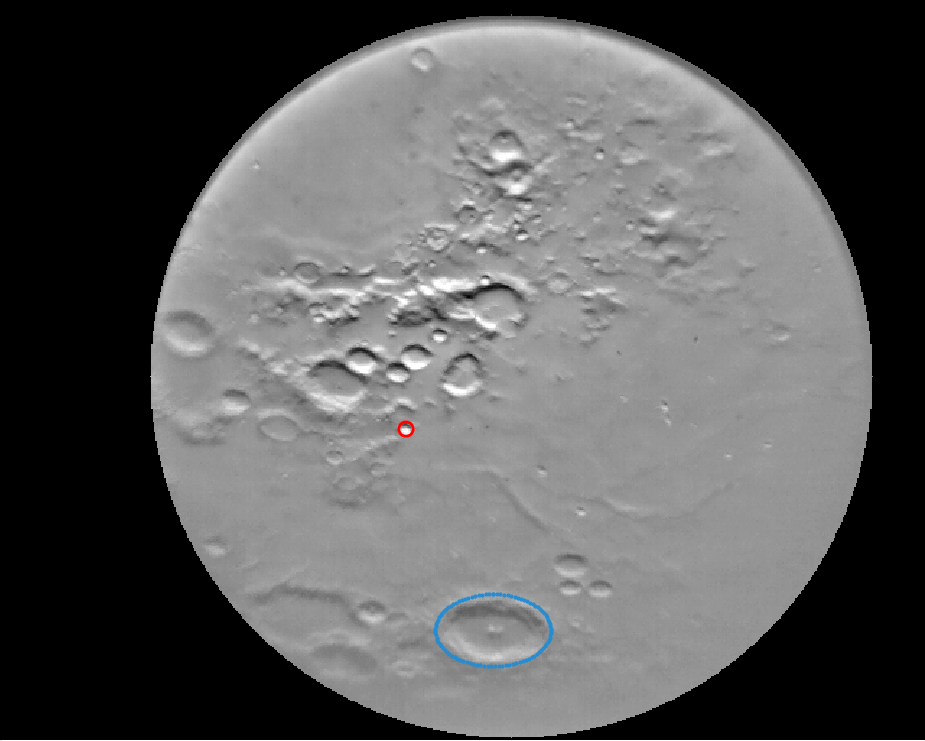}
        }\\%
        \subfigure[Edge during penumbral eclipse]{%
            \label{fig:fifth}
            \includegraphics[width=0.4\textwidth]{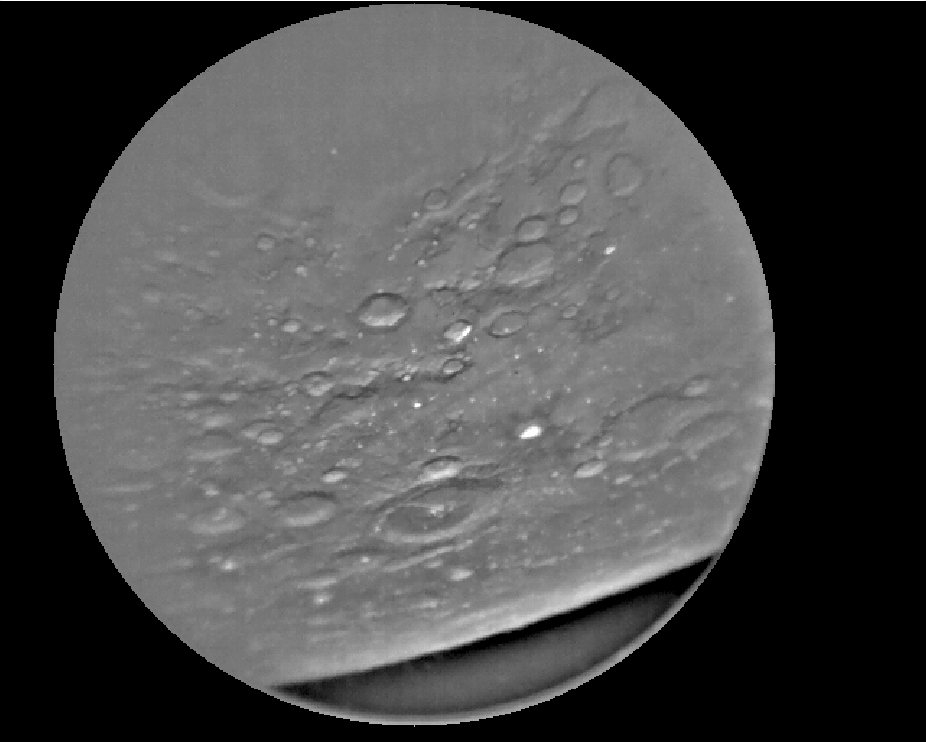}
        }%
        \subfigure[Edge at Maximum eclipse]{%
            \label{fig:sixth}
            \includegraphics[width=0.4\textwidth]{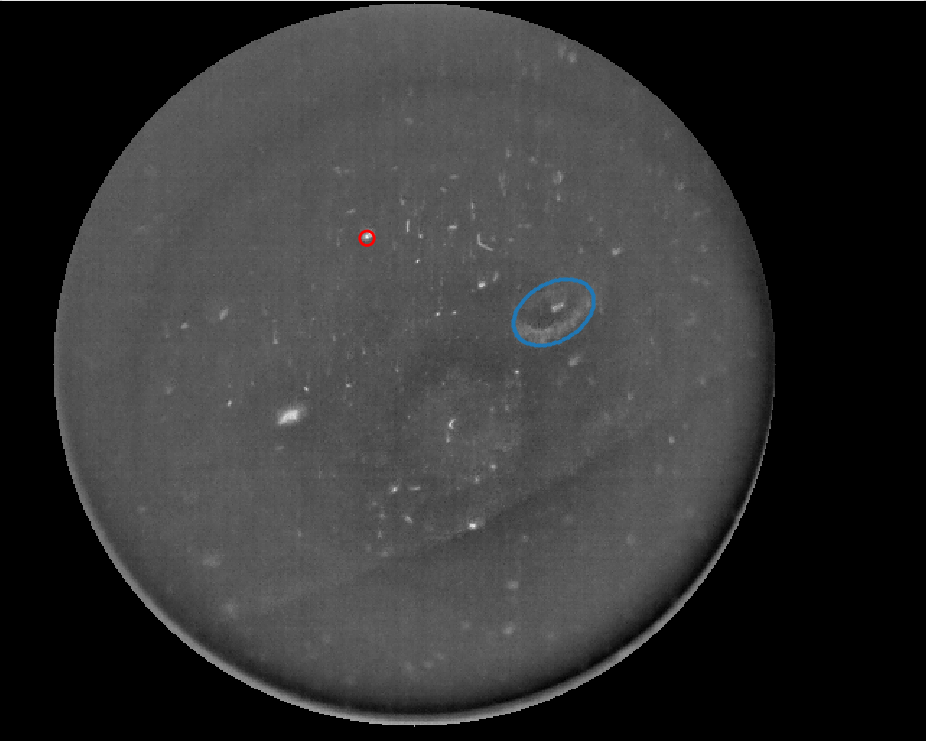}
        }%
    \end{center}
    \caption{%
   Observations of the moon during the lunar eclipse and follow-up on 2019 Jan 22. {Lunar limb} observations highlight the change in temperature between early and maximum eclipse. Where possible, the locations of two craters: Bellot (red) and Langrenus (blue), have been indicated. {The presence of dust on the field lens can masquerade as bright features on the lunar surface, this can be seen very clearly in image (f) where bright features lie beyond the limb. As these dust contaminants remain fixed in position in the FOV, genuine bright features have been confirmed through their movement with the lunar surface.} For scale, one pixel $\sim$ 4 km on the lunar surface.
     }%
   \label{fig:eclipse}
\end{figure*}
\begin{figure*}
    \centering
    \includegraphics[width=\textwidth]{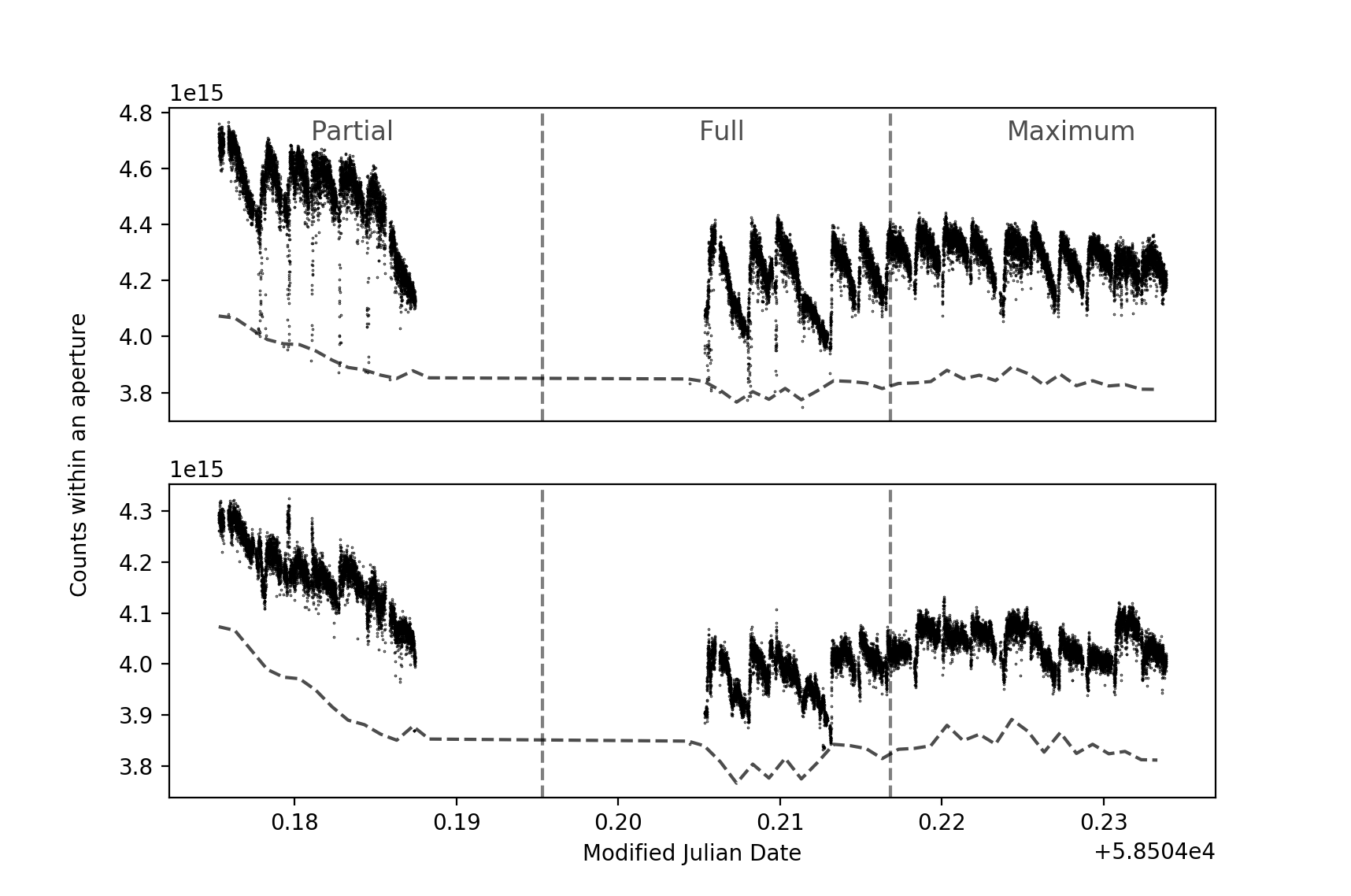}
    \caption{Observations of Bellot (top) and Langrenus (bottom) during the lunar eclipse on 2019 Jan 21. {Count} values are maximum values within an aperture. Both curves are plotted with background {counts} (dashed line). A large period between partial and full eclipse went unobserved due to a sudden spike in humidity that halted observation. The periodic fluctuating pattern present in all curves is thermal drift. This is described in more detail in section \protect\ref{calibration}.}
    \label{fig:heatloss}
\end{figure*}

Observations of the Jan 21, 2019 eclipse and full moon ($\pm$ 2 days) were taken to obtain measurements of surface temperature and its variability during an eclipse. The eclipsing lunar disk has been well studied since the first observations of thermal anomalies in 1960 (\citealt{Short60}; \citealt{SAA63}; \citealt{Short65}; \citealt{SAA66}; \citealt{FUD66}; \citealt{Hunt68}; \citealt{Short72}; \citealt{WIN72}; \citealt{Foun76}; \citealt{Pri03}; \citealt{Law03}).

The general thermophysical properties of the lunar surface have been well mapped in the mid-IR (see \citealt{Pai10}; \citealt{Vasa12} and references therein). The low thermal conductivity of the fine-grained, regolith that makes up the upper $\sim$ 0.02m surface layer, results in extreme diurnal temperature variation. During lunar daytime, illuminated surfaces are close to radiative equilibrium and high surface temperatures (approaching 400K) are a result of incident solar flux. Comparatively, during a lunar eclipse or during lunar nighttime, with no incident solar radiation and no atmosphere to trap heat, surface temperatures drop by $\sim$ 300K. The moon also shows significant topographic heterogenity. This results in high spatial variation in the thermophysical properties of lunar surface features.

A simple temperature model for the lunar surface (equation \ref{eq:est_temp}), derived by \cite{Shaw15}, can be used to estimate the expected temperature of the lunar surface, $T_{m}$ at any given time; where $E_{m}$ is the spectrally averaged solar irradiance at the surface (adjusted for seasonal variations in earth sun distance). The subsolar point, where lunar latitude angle $\varphi=0$, can be used to approximate the maximum temperature of the lunar surface, $T_{max}$. Latitudinal and diurnal variation in $T_{m}$ across increasing lunar radius is mostly controlled by angular distance to the subsolar point. 
\begin{equation}
    T_{m}(\varphi) = \bigg(\frac{E_{m}\ \cos(\varphi)}{\sigma}\bigg)^{1/4}=\ T_{max}\ \cos^{1/4}(\varphi)
    \label{eq:est_temp}
\end{equation}
Using equation \ref{eq:est_temp}, we can calculate that the expected maximum temperature of the lunar surface on the 19th and 22nd of January 2019, two days either side of the full moon, is 396.8K. The radiation incident on the Liverpool telescope can be modelled as:
\begin{equation}
    I(\phi_{z},\lambda)\ =\ I_{0}\ e^{-AM(\phi_{z})\cdot \tau_{N}(\lambda)}
    \label{eq:transmission}
\end{equation}
where $AM(\phi_{z})$ is the airmass for a given zenith angle and $\tau_{N}(\lambda)$ is the normal optical thickness. $\tau_{N}(\lambda)$ is dependent on the atmospheric transmission for a given wavelength. For mid-IR wavelengths, the theoretical atmospheric transmission at normal incidence ranges from 0.70 to 0.75 \citep{Vol12}. Equation \ref{eq:transmission} therefore gives an expected value of observed lunar surface temperature of 362.9K. Our observations on these dates recorded an average temperature of 350 $\pm$ 6 K. This slightly lower value is likely due to losses in the telescope and instrument optics.

Observations of partial, full and maximum phases of the lunar eclipse were taken (see figure \ref{fig:eclipse}). Significant heat loss occurs prior to this, during the penumbral phase \citep{Vol12}. Unfortunately, we were unable to obtain enough sequential observations of the same region due to poor weather conditions to include these exposures in our analysis.

Visually, the change in lunar surface from partial eclipse onwards is quite significant. As the eclipse reaches totality, the different thermophysical properties of features of the lunar surface become very apparent. During this time, many features become unobservable. However, the hundreds of thermal anomalies that were first seen by \cite{Short60} appear as very bright `hot spots' during partial eclipse and remained bright throughout totality. During our observations, we focused on several lunar features in the mare Fecunditatis region (7.8$^\circ$S, 51.3$^\circ$E). Analysis was conducted for two craters; Bellot (12.49$^\circ$S, 48.2$^\circ$E) and Langrenus (9$^\circ$S, 62$^\circ$E). Langrenus is an early Copernican crater with a faint ray pattern, a high albedo, a central peak of brecciated bedrock and a moderate thermal anomaly \citep{Short72}. Bellot is a smooth, dark haloed crater with a high albedo, likely a result of freshly exposed, brecciated rock \citep{Els67}, resulting in a significant thermal anomaly.

To conduct data analysis for $\sim$ 30000 eclipse exposures we employed a semi-automated feature tracking regime to determine the approximate centre of the brightest source, Bellot. The coordinates of Bellot were then used to anchor the movement of other features in the field of view. Each exposure was reduced, and raw values converted to {counts}, as described in section \ref{calibration}. {Counts} for each feature were obtained by taking the maximum value from apertures containing each feature. A circular aperture with 5 pixel diameter was used for Bellot and an elliptical aperture with 30 pixel semi-major axis, and $\pi /3$ rotation for Langrenus.

Analysis of observations during partial eclipse showed approximate temperature loss rates of 0.98K per minute for Bellot and 0.50K per minute for Langrenus. Comparatively the average heat loss in regions with no `hot spots' or crater features occurred at a rate of 0.26K per minute. Figure \ref{fig:heatloss} are {count} measurements of Bellot (top) and Langrenus (bottom) during the eclipse. The dashed line in both plots represents the background {counts} recorded from an aperture with 10 pixel diameter, in a region of mare with no crater features. The apparent plateau in all three curves is likely a result of telescope and system emissions limiting the range of temperatures that can be recorded.

The low conductivity of the upper lunar regolith results in little exchange of energy between warmer subsurface and surface layers. As a result, the lunar surface cannot maintain surface temperatures without incident radiation from the Sun. Couple this with the high emissivity of the lunar surface around full moon ($\epsilon$ $\sim 0.97$ \citealt{Shaw15}), we can approximate that the energy required $\Delta E$, to maintain the lunar daytime temperature $T$, is equal to the energy release over time $dt$, during eclipse:
\begin{equation}
    \Delta E = mc \Delta T dt= \epsilon \sigma T^{4}
    \label{eq:rad_eq}
\end{equation}
where $m$ is mass, $c$ is the specific heat capacity and $\sigma$ is the Stefan-Boltzmann constant. We can derive the following differential equations from equation \ref{eq:rad_eq} for a temperature change of $T_{1}$-$T_{2}$ in time $t_{1}$ to $t_{2}$:
\begin{equation}
    \int_{T_{1}}^{T_{2}} \frac{1}{T^{4}}\ dT = \int_{t_{1}}^{t_{2}} \frac{\epsilon \sigma}{mc}\ dt
    \label{eq:diff}
\end{equation}
The bulk density of the lunar regolith sharply increases at a depth of $\sim 0.02$m. So, solving equation \ref{eq:diff} for a 1m$^{2}$ area, we can calculate a naive estimate of specific heat capacity $c$, for different regions in our observations. For Bellot and Langrenus, we calculate a specific heat capacity of 2.3 and 4.1 kJ/kg/K/ respectively. There is a discrepancy between our naive estimates and the heat capacity of the lunar regolith quoted in literature. Analysis of the Apollo 14, 15 and 16 samples recorded specific heat capacities of between 0.21 and 0.8 kJ/kg/K \citep{Hem73}. More recent studies of sintered Australian Lunar Regolith Simulant (ALRS-1) have found values of up to 1.63 kJ/kg/K \citep{Bon14}. The disgreement between these values and our own could be largely in part due to the Apollo samples being collected from regions different than ours. This would be applicable to the values obtained from ALRS-1 as it is created to have a chemical composition comparative to Apollo 12 samples. We also make several assumptions in our calculations that may not be applicable to our crater regions. The quoted emissivity and depth values are approximated across the entire lunar surface, but these are known to vary with region age and regolith material. It is also possible that the shape of both craters contributes to the storage of heat during eclipse.

\section{Discussion/conclusion}
In this paper we have presented a prototype instrument that adapts mid-IR uncooled microbolometer technology for use on ground telescopes in the 1-2m class. For this purpose, additional optics were designed to rescale the image onto the detector to optimally sample the diffraction limit. We opted to design and build the instrument from commercially available units, at a low cost. The instrument was tested on the LT over 3 days in 2019 Jan. A small programme of observations of solar system and stellar objects was conducted. From these observations we confirmed a plate scale of 0.75$^{\prime\prime}$ per pixel and obtained a measure of the $\sim$ 10$\%$ photometric stability and performance of the instrument. We recorded a 3$\sigma$ sensitivity of $\sim 7 \times 10^{3}$ Jy for a single exposure corresponding to a sensitivity limit of $\sim 3 \times 10^{2}$ Jy for an integration time of 60 seconds. Using the IRAS point source catalog v2.1 \citep{iras98} we can see that such a limit would make a further $\sim$ 163 extra-solar sources observable with the current instrument setup. {Given the IRAS point spread function at 12$\mu m$ corresponds to a FWHM of < 16$^{\prime \prime}$ some of these sources may be extended and therefore our calculation is an estimate of the upper limit of observable sources.} For observations of bright asteroids, such as Ceres, this limit would have to be improved by a factor of two \citep{Mul02}. 
Observations of the eclipsing moon are presented as a science case. In general, the overwhelming sky and telescopic emission limited observations to very bright mid-IR sources and made data reduction difficult. {The germanium foreoptics have significant transmission into the wings of the N-band. As a result, there is likely excess sky noise limiting our observations which could be improved with the inclusion of a narrow bandpass filter.} We aim to further develop the prototype to include a chopping/nodding regime to test whether the stability and sensitivity of the system can be improved and increase the number of observable sources.

\section*{Acknowledgements}
MFR and IAS gratefully acknowledge funding from the Royal Astronomical Society Patricia Tomkins Instrumentation award. The LT is operated by Liverpool John Moores University at the Spanish Observatoria del Roque de los Muchachos of the Instituto de Astrof\'{i}sica de Canarias in La Palma, with financial support from the UK Science and Technology Facilities Council. MFR is funded by a Liverpool John Moores University FET/SCS PhD scholarship. This research has made use of the NASA/ IPAC Infrared Science Archive, which is operated by the Jet Propulsion Laboratory, California Institute of Technology, under contract with the National Aeronautics and Space Administration. This research made use of Astropy, a community-developed core Python package for Astronomy \citep{astropy:2013, astropy:2018}




\bibliographystyle{mnras}
\bibliography{example}







\bsp	
\label{lastpage}
\end{document}